\documentclass[11pt,a4paper]{article}

\usepackage{geometry}
\usepackage[applemac]{ inputenc}

\usepackage{mathrsfs}
\usepackage{amsmath}
\usepackage{amssymb}
\usepackage{multicol}
\usepackage{float}
\usepackage{makeidx}
\usepackage{layout}
\usepackage{array}
\usepackage{boxedminipage}
\usepackage{hyperref}
\usepackage{latexsym}
\usepackage{indent first}

\usepackage{subcaption,graphicx}

\usepackage{color}

\usepackage[sectionbib,square]{natbib}

\usepackage[normalem]{ulem}

\title{An iterative method to compute conformal mappings and their inverses in the context of water waves over topographies}
\author{M. V. Flamarion$^{1}$ and R. Ribeiro-Jr$^{2}$}

\date{}

\begin{document}

\maketitle

{\footnotesize
	\begin{center}
        $^{1}$ UFRPE/Rural Federal University of Pernambuco, UACSA/Unidade Acad{\^e}mica de Cabo de Santo Agostinho, BR 101 Sul, 5225, 54503-900, Ponte dos Carvalhos, Cabo de Santo Agostinho, Pernambuco, Brazil.

	$^2$ UFPR/Federal University of Paran\'a,  Departamento de Matem\'atica, Centro Polit\'ecnico, Jardim das Am\'ericas, Caixa Postal 19081, Curitiba, PR, 81531-980, Brazil.

	\end{center}
	
	}

\begin{abstract}

An iterative numerical method to compute the conformal mapping in the context of propagating water waves over uneven topographies is investigated. 
The map flattens the fluid domain onto a canonical strip in which computations are performed. 
The accuracy of the method is tested by using the MATLAB Schwarz-Christoffel toolbox mapping as a benchmark. 
 Besides, we give a numerical alternative to compute the inverse of the  conformal map. \\


\end{abstract}

\section{Introduction}
The Euler equations are the main model used to study hydrodynamic problems. For instance, flow of water over rocks (\cite{Pratt}), atmospheric flows encountering obstacles (\cite{Baines}), ship wakes (\cite{Grimshaw}) and waves generated by storms (\cite{Johnson}). Solving Euler equations numerically is a hard problem due to its nonlinearity and its boundary conditions. One of the main tools used to solve numerically the two-dimensional Euler equations associate with water waves is the conformal mapping technique. 

The conformal mapping technique  in the context of water waves was initially introduced by \cite{Dyachenko}. This approach is used to transform a free boundary problem given by Euler equations into a family of ordinary differential equations which are easier to be solved numerically. 
We summarize here some works which in this method has been used.

Studying water waves propagating in a channel with flat bottom,  \cite{Choi99} used the  conformal mapping technique  to flatten the free surface and map the fluid domain onto a  strip. This allowed them to compare solitary wave solutions of Euler equations with the ones produced by the Korteweg-de Vries equation (KdV). 
 Later, \cite{Choi09} obtained traveling waves for Euler equations in a sheared channel with constant vorticity and flat bottom. In this scenario, particle trajectories and the pressure within the bulk of the fluid were computed in \cite{JFM17}. \cite{Milewski2} used a conformal mapping to study capillary-gravity waves in a channel with infinity depth. In this work,  the semiplane fluid domain was mapped onto an infinity strip and the numerical stability of traveling waves and collisions were investigated. 
 
Investigating waves over an uneven topography, \cite{Terreno} constructed a conformal mapping to flatten the bottom topography and computed numerical solutions for the Euler equations. Using asymptotic analysis  he obtained a Boussinesq-type equation with variable coefficients. In this work the conformal mapping was computed using the MATLAB Schwarz-Christoffel toolbox (SC--toolbox). Beyond the free-surface problem the conformal mapping technique has also been used in the study of internal waves. For instance, for a two-layer fluid of finite depth, \cite{Zarate} used the conformal mapping technique to derive a higher-order nonlinear model to study the interaction of nonlinear internal waves with large amplitude bottom topography.

Analysing the behavior of waves generated due to a current-topography interaction, \cite{Flamarion} constructed a conformal mapping to flatten simultaneously the free surface and the topography.
They assumed that the topography had small amplitude and validated their
numerical method comparing solutions of the Euler equations in the weakly nonlinear and weakly dispersive regime with solutions of the Forced Korteweg-de Vries (fKdV) equation. More recently,
\cite{FRN} used the same technique to study trajectories of particles in a finite depth sheared channel in the presence of a variable topography with small amplitude for the linear Euler equations. They  found stagnation points in the fluid domain and the formation  of a time-dependent Kelvin cat-eye structure.

In this work, we study an iterative method to compute the conformal mapping that maps the region $\Omega=\{(x,y)\in\mathbb{R}^{2}; -L\le x\le L, \;\ \mbox{and} \;\ -1+h(x)\le y\le0\}$ onto a strip. In the context of water waves, this region can be read as a fluid domain, in which $y=0$ is the free surface and $y= -1+h(x)$ is the bottom topography, where $h$ is a smooth function. 
 The novelty in this paper is that we consider topographies ($h$) with large amplitudes. The numerical method is validated by comparing it with  the MATLAB SC--toolbox. Furthermore, we present a new alternative to compute the inverse of the conformal mapping.

This article is organized as follows. In section 2, we formulate the conformal mapping. The numerical results are presented in section 3 and the conclusion in section 4.

\section{Mathematical Formulation}
\begin{figure}[h!]
\centerline{\includegraphics[width=0.9\textwidth]{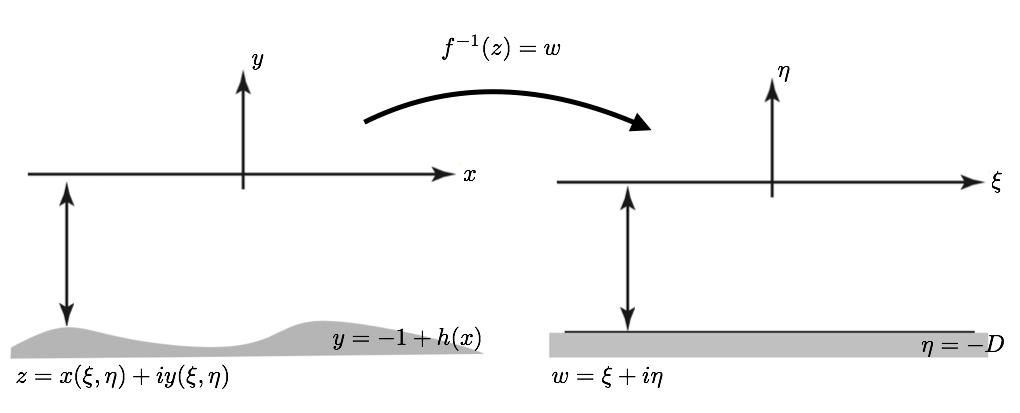}}
\caption{The inverse conformal mapping. The topography is flattened out in the canonical domain.}
\label{Conforme}
\end{figure}

Consider the region $\Omega=\{(x,y)\in\mathbb{R}^{2}; -L\le x\le L, \;\ \mbox{and} \;\ -1+h(x)\le y\le0\}$ on the $xy$-plane, where $h$ is a periodic smooth function with period $2L$.  We construct a conformal mapping $f$ 
\begin{equation}\label{direta}
f(\xi+i\eta) = x(\xi,\eta)+iy(\xi,\eta),
\end{equation}
which flattens the bottom topography  onto a strip of height $D$. The conformal mapping satisfyies the boundary conditions 
\begin{equation*}
y(\xi,0)=0 \;\ \mbox{and} \;\ y(\xi,-D)=-1+\mathbf{H}(\xi),
\end{equation*}
where $\mathbf{H}(\xi)=h(x(\xi,-D))$. The inverse of the conformal mapping is 
\begin{equation}\label{inversa}
f^{-1}(x+iy)= \xi(x,y)+i\eta(x,y),
\end{equation}
and its sketch is depicted in Figure \ref{Conforme}.

The functions $x(\xi,\eta)$ and $y(\xi,\eta)$ in (\ref{direta})  are harmonic conjugate. Therefore, due to the boundary condition we have 
\begin{align}\label{Laplace1}
\begin{split}
& y_{\xi\xi}+y_{\eta\eta}=0 \;\ \mbox{for} \;\ -D < \eta < 0, \\
& y = 0,\;\  \mbox{at} \;\ \eta =0, \\
& y = -1+\mathbf{H}(\xi),\;\  \mbox{at}\;\ \eta =-D. \\
\end{split}
\end{align}
The Dirichlet boundary conditions express the bottom corrugations in the domain. As the problem (\ref{Laplace1}) is linear we can express its solutions in terms of Fourier series as
\begin{align}\label{A2}
\begin{split}
& y(\xi,\eta) = \mathcal{F}^{-1}_{k_j\ne 0}\bigg[\frac{-\coth(k_jD)\sinh(k_j\eta)\widehat{\mathbf{H}}}{\cosh(k_jD)}\bigg] + \frac{1-\widehat{\mathbf{H}}(0)}{D}\eta.\\
\end{split}
\end{align}
Fourier modes are given by $$\mathcal{F}_{k_j}[g(\xi)]=\hat{g}(k_j)=\frac{1}{2L}\int_{-L}^{L}g(\xi)e^{-ik_j\xi}\,d\xi,$$
$$\mathcal{F}^{-1}_{k_j}[\hat{g}(k_j)](\xi)=g(\xi)=\sum_{j=-\infty}^{\infty}\hat{g}(k_j)e^{ik_j\xi},$$
where $k_j=(\pi/L)j$, $j\in\mathbb{Z}$. The Cauchy-Riemann equation $x_\xi=y_\eta$ yields
\begin{align}\label{xx}
\begin{split}
& x(\xi,\eta) = \mathcal{F}^{-1}_{k_j\ne 0}\bigg[\frac{i\coth(k_jD)\cosh(k_j\eta)\widehat{\mathbf{H}}}{\cosh(k_jD)}\bigg] + \frac{1-\widehat{\mathbf{H}}(0)}{D}\xi.\\
\end{split}
\end{align}

Let $\mathbf{X}(\xi)$ be  the horizontal  coordinate at $\eta=0$ and $\mathbf{X}_{b}(\xi)$ the respective trace along the bottom $\eta=-D$. From (\ref{xx}) we have
\begin{align}\label{xxi}
\begin{split}
& \mathbf{X}(\xi) = \frac{1-\widehat{\mathbf{H}}(0)}{D}\xi+ \mathcal{F}^{-1}_{k_j\ne 0}\bigg[\frac{i\coth(k_jD)\widehat{\mathbf{H}}}{\cosh(k_jD)}\bigg], \\
& \mathbf{X}_{b}(\xi)=x(\xi,-D)=\frac{1-\widehat{\mathbf{H}}(0)}{D}\xi+ \mathcal{F}^{-1}_{k_j\ne 0}\bigg[i\coth(k_jD)\cosh(k_j\eta)\widehat{\mathbf{H}}\bigg].  
\end{split}
\end{align}
It is important to notice that in the equation $(\ref{xxi})_{2}$, $\mathbf{X}_{b}(\xi)$ is defined implicitly. 
We impose that $\mathbf{H}(\xi)$ to be $2L$-periodic in the $\xi$-variable.  Since the conformal mapping preserves angles and the horizontal length scale remains unchanged,
the width of the canonical strip has to adjust accordingly. From $(\ref{xxi})_{1}$ we obtain
$$\big<\mathbf{X}_{\xi}\big>=\frac{1-\big<\mathbf{H}\big>}{D}.$$
Hence, we choose $D=1-\big<\mathbf{H}\big>$. So we have the following equations
\begin{align}\label{xxf}
\begin{split}
& \mathbf{X}(\xi) = \xi+ \mathcal{F}^{-1}_{k_j\ne 0}\bigg[\frac{i\coth(k_jD)\widehat{\mathbf{H}}}{\cosh(k_jD)}\bigg],\\
& \mathbf{X}_{b}(\xi)=\xi+ \mathcal{F}^{-1}_{k_j\ne 0}\bigg[i\coth(k_jD)\cosh(k_j\eta)\widehat{\mathbf{H}}\bigg].  
\end{split}
\end{align}

In the next section we present a numerical method to solve (\ref{xxf}) and compare the results with the SC--toolbox. Besides, we present a numerical method to compute the map $f^{-1}$ given in (\ref{inversa}).

\section{Numerical methods and results}

In this section we construct an iterative method to solve (\ref{xxf}).
Fourier modes are computed by the fast Fourier Transform (FFT) 
on a uniform grid. Derivatives in the $\xi$--variable are performed spectrally (\cite{Trefethen}). 

The equation $(\ref{xxf})_{2}$ shows that $\mathbf{X}_{b}(\xi)$ and $\mathbf{H}(\xi)$ are coupled in a nontrivial fashion. 
We solve it using the following iterative scheme:
\begin{align}\label{scheme}
\begin{split}
& \mathbf{X}_{b}^{l}(\xi)=\xi+ \mathcal{F}^{-1}_{k_j\ne 0}\bigg[i\coth(k_jD)\cosh(k_j\eta)\widehat{\mathbf{H}^{l}}\bigg],  \\
& \mathbf{H}^{l+1}(\xi)=h(\mathbf{X}_{b}^{l}(\xi)).
\end{split}
\end{align}
The initial step is  $\mathbf{X}_{b}^{0}(\xi)=\xi$ and $\mathbf{H}^{1}(\xi)=h(\xi)$. The scheme is performed with  the stopping criteria
$$\frac{\displaystyle\max_{\xi\in[-L,L]}\Big|\mathbf{H}^{l+1}(\xi)-\mathbf{H}^{l}(\xi)\Big|}{\displaystyle\max_{\xi\in[-L,L]}\Big|\mathbf{H}^{l}(\xi)\Big|}<\epsilon,$$
where $\epsilon>0$ is a given tolerance. Once $\mathbf{H}(\xi)$  is computed through (\ref{scheme}), we can easily obtain $\mathbf{X}(\xi)$ from the formula  $(\ref{xxf})_{1}$.
\subsection{Benchmarking}

In the following simulations we consider that the bottom topography is modelled by the function $$h(x)=Ae^{-\lambda x^2}.$$
Since $h(x)$ decays exponentially to zero as $|x|\rightarrow\infty$, we can truncate its domain to $-L<x<L$, and approximate the boundary conditions  by periodic conditions.
The parameters used in all simulations are $L = 15$, $N = 2^{10}$, $\Delta\xi=2L/N$, $\lambda>0$ and the tolerance is $\epsilon=10^{-16}$. 

\begin{figure}[h!]
\centerline{\includegraphics[width=0.9\textwidth]{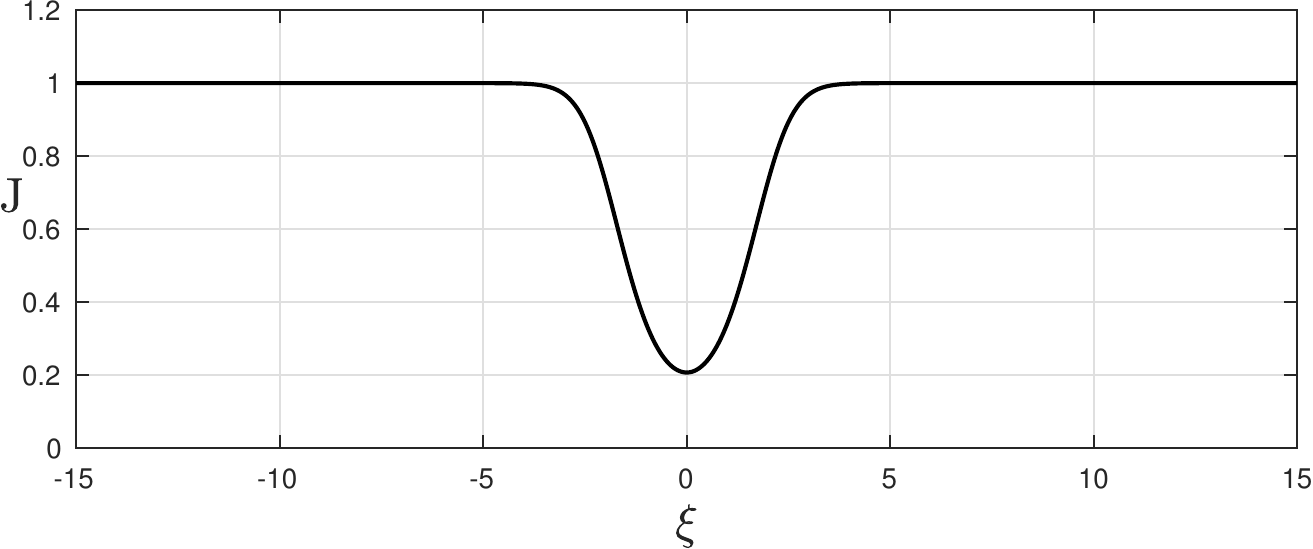}}
\caption{The Jacobian computed using the iterative method for $A=0.5$.}
\label{JacA05}
\end{figure}

Once the application $G(z)=\sqrt{\lambda} z$, $z\in\mathbb{C}$, is conformal the composition $G\circ f^{-1}$ is conformal as well. Thus, without loss of generality we can assume that $\lambda=1$. Let $J=\mathbf{X}_{\xi}^2(\xi)$ be the  mapping's Jacobian of the iterative
method evaluated at $\eta=0$ and $J_{sc}$ the  mapping's Jacobian evaluated at $\eta=0$ computed using the SC--toolbox. 
Figure \ref{JacA05} depicts the Jacobian $J$, which is computed  for $A=0.5$. We notice that when we move away from where the topography is, the Jacobian is effectively equal to 1. We verify the accuracy
of our numerical method comparing  $J$ and $J_{sc}$. 

Table \ref{table1} displays the absolute and relative errors of the Jacobians $J$ and $J_{sc}$ for different values of $A$. They agree very well. Although the error increases as the amplitude of the topography increases, in the worst case scenario ($A=0.5$) both the absolute and relative errors are still of order $\mathcal{O}(10^{-5})$. However, this is not a limitation of the method, because  in water wave problems we usually consider obstacles with amplitude $A<0.5$, which physically means that the amplitude is $50\%$ of the depth channel. 

In order to measure the computational effort required to compute the conformal mappings via the iterative method and using the SC--toolbox, we monitored the computational time spent to compute the Jacobians $J$ and $J_{sc}$ through the command {\it tic toc} on MATLAB. The iterative method proved to be very efficient. While $J$ was computed in approximately 0.1 seconds,  $J_{sc}$ was computed roughly in 5000 seconds.
In water waves problems, the conformal mapping has to be computed every timestep, so based on our results the iterative method is more suitable to study these type of problems.
\begin{table}[h!]
\centering
\begin{tabular}{c|c|c}\hline\hline
$A$ & $\displaystyle\max_{\xi \in [-L,L]}{\Big|J-J_{sc}\Big|}$ &  $\frac{\displaystyle\max_{\xi \in [-L,L]}{\Big|J-J_{sc}\Big|}}{\displaystyle\max_{\xi \in [-L,L]}{\Big|J_{sc}\Big|}}$  \\   \hline
0.01 	 &	 1.70$\times 10^{-7}$&  1.70$\times 10^{-7}$ \\ \hline
0.03	 &	4.42$\times 10^{-7}$&	 4.40$\times 10^{-7}$ \\ \hline
0.05	 &	8.88$\times 10^{-7}$ &  8.82$\times 10^{-7}$\\ \hline 
0.10	  &	  1.85$\times 10^{-6}$ & 1.82$\times 10^{-6}$ \\ \hline 
0.20	 &	4.00$\times 10^{-6}$ &	3.89$\times 10^{-6}$\\ \hline
0.30	 	& 6.19$\times 10^{-6}$&  5.90$\times 10^{-6}$  \\ \hline
0.40  	 &	 8.36$\times 10^{-6}$& 	7.79$\times 10^{-6}$\\ \hline 
0.50  	 & 	1.19$\times 10^{-5}$ & 	1.07$\times 10^{-5}$\\  \hline
\end{tabular}
\caption{The absolute and relative error of the iterative Jacobian map.}\label{table1}
\end{table}

Lastly, we test the resolution of the iterative method. To do that, we set $J_{N_{0}}$ as reference solution computed on our finest grid, with $N=N_{0}=4096$ and fix $A=0.5$. Recall that in this article we considered $N=1024$. Table \ref{table2} displays the absolute and relative error in the   $\ell_2$-norm. It is clear that the Jacobian does not depend on the grid. 
\begin{table}[h!]
\centering
\begin{tabular}{c|c|c}\hline\hline
$N$ & $\displaystyle{\Big|\Big|J_{N_{0}}-J\Big|\Big|_{2}}$ &  $\displaystyle\frac{\Big|\Big|J_{N_{0}}-J\Big|\Big|_{2}}{\Big|\Big|J_{N_{0}}\Big|\Big|_{2}}$  \\   \hline
128 	 &	 9.29$\times 10^{-10}$&  7.90$\times 10^{-11}$ \\ \hline
256	 &	1.90$\times 10^{-15}$ &	1.14$\times 10^{-16}$ \\ \hline
512	 &	2.53$\times 10^{-15}$& 1.07$\times 10^{-16}$\\ \hline 
1024	  &	 3.61$\times 10^{-15}$	& 1.08$\times 10^{-16}$ \\ \hline 
2048	 &	5.21$\times 10^{-15}$ &	1.10$\times 10^{-16}$ \\ \hline
\end{tabular}
\caption{The absolute and relative error of the iterative Jacobian map computed in different grids.}\label{table2}
\end{table}

The iterative conformal mapping can be promptly applied to solve the linear Euler equations. For the full nonlinear Euler equations, in which the upper boundary of the fluid domain is a time-dependent function, the method presented here can easily be extended. For this purpose, one can compute the conformal mapping at every timestep in the same fashion as showed here.  Since 
 problem (\ref{Laplace1}) is linear, the formulas of the new conformal mapping will be the same as in (\ref{A2})-(\ref{xx}) plus a term which only depends on the free surface.

\subsection{Inverse conformal mapping}
In this section we present a method based on an optimization problem to compute the inverse conformal mapping given in (\ref{inversa}) through the formulas (\ref{A2})-(\ref{xx}), i.e., given $(x_{0},y_{0})$  in the physical domain $\Omega$
we find $(\xi_0,\eta_0)$ in the canonical domain such as $$x(\xi_0,\eta_0)=x_0,\;\ \mbox{and}\;\ y(\xi_0,\eta_0)=y_0.$$
In order to do that we set the following minimization problem:
\begin{equation}
\begin{aligned}
& \min \quad \psi(\xi,\eta)\\
& \text{s.t. } -L\leq\xi\leq L\\
&\;\ \;\ \;\ -D\leq\eta\leq 0
\end{aligned}\label{opt-P}
\end{equation}
where $\psi(\xi,\eta)= (x(\xi,\eta)-x_{0})^{2}+ (y(\xi,\eta)-y_{0})^{2}$. Notice that the conformal mapping is one-to-one, so  the problem (\ref{opt-P}) always has an unique solution for any choice of $(x_{0},y_{0})\in\Omega$. Thus, we solved the optimization problem (\ref{opt-P}) using the MATLAB Optimization toolbox obtaining $f^{-1}(x_0,y_0)$. 

To verify the accuracy of this method, the inverse of confomal mapping (\ref{A2})-(\ref{xx}) is computed in the discrete domain $$\Omega_{d}=\{(x_i,y_j); \;\ x_i=-L+(i-1)\Delta x, \;\  y_j=-1+h(x_i)+(j-1)\Delta y^{i}\},$$ 
where $\Delta x =2L/N$ and  $\Delta y^{i} = (1-h(x_i))/M$, for $i=1,2,\dots ,N$, $j=1,2,\dots ,M+1$. Denoting by $f_d$ the conformal mapping computed numerically and by $f^{-1}_d$
its inverse computed as mentioned above, we compare $f_{d}\circ f_{d}^{-1}$ with the identity function. To this end, we define the error as
$$ Error = \frac{\displaystyle\max_{\Omega_{d}}||f_{d}\circ f_{d}^{-1}(x_i,y_j)-(x_i,y_j)||_{\infty}}{\max||(x_i,y_j)||_{\infty}}.$$
The results displayed on Table \ref{table3} show that the method has high accuracy. 
\begin{table}[h!]
\centering
\begin{tabular}{c|c}\hline\hline
$A$ &  $Error$  \\   \hline
0.10	 & $3.16\times 10^{-8}$ \\ \hline 
0.20	 &	$3.20\times 10^{-8}$\\ \hline
0.30	 &  $3.60\times 10^{-8}$  \\ \hline
0.40  & 	$3.37\times 10^{-8}$ \\ \hline 
0.50  & 	$4.28\times 10^{-8}$\\  \hline
\end{tabular}
\caption{The error obtained in the inverse conformal mapping (\ref{A2})-(\ref{xx}) computation. Parameters: $N=1024$ and $M=10$.}\label{table3}
\end{table}

It is worth pointing out that the methodology presented here is computationally easy to implement and as discussed in the previous section it requires low computational time.

\section{Conclusion}
 In this paper, we have studied an iterative method to compute a conformal mapping which can be used in water waves problems with topographic obstacles. We used the SC--toolbox as a benchmark to validade the numerical method. The method turned out to be of easy implementation and with low computational time presenting itself as an alternative to compute conformal mappings. Although we could notice that the error increases as we increase the amplitude of the topography, the accuracy is still very good for topographies with amplitudes lesser than $50\%$ of the depth. Besides, we presented a numerical method to compute the inverse of the iterative conformal mapping. 
 
 \section{Acknowledgements}

The authors are grateful to IMPA-National Institute of Pure and Applied Mathematics for the research support provided during the Summer Program of 2020.
M.F.   is grateful to Federal University of Paran{\' a} for the visit to the Department of Mathematical Sciences. 
R.R.-Jr  is grateful to University of Bath for the extended visit to the Department of Mathematical Sciences.

\bibliographystyle{abbrv}

\end{document}